\documentclass[aps, twocolumn,amsmath,amssymb]{revtex4}

\usepackage{hyperref}
\usepackage{graphicx}
\usepackage{dcolumn}
\usepackage{xcolor}
\usepackage{braket}
\usepackage{commath}
\usepackage{adjustbox}
\usepackage{comment} 
\usepackage{mathtools}

\newcommand{\bs}[1]{{\boldsymbol{#1}}}

\newcommand{\bq}{\bs{q}}

\newcommand{\br}{\bs{r}}

\begin{document}

\title{Classical Casimir force from a quasi-condensate of light}

\author{Tamara Bardon-brun}
\affiliation{Laboratoire Kastler Brossel, Sorbonne Universit\'e, CNRS, ENS-PSL University, Coll\`ege de France; 4 Place Jussieu, 75005 Paris, France}

\author{Simon Pigeon}
\affiliation{Laboratoire Kastler Brossel, Sorbonne Universit\'e, CNRS, ENS-PSL University, Coll\`ege de France; 4 Place Jussieu, 75005 Paris, France}

\author{Nicolas Cherroret}
\email{cherroret@lkb.upmc.fr}
\affiliation{Laboratoire Kastler Brossel, Sorbonne Universit\'e, CNRS, ENS-PSL University, Coll\`ege de France; 4 Place Jussieu, 75005 Paris, France}

\begin{abstract}
We show that weakly incoherent optical beams propagating in a Kerr medium exhibit a universal algebraic coherence after a short propagation time, mimicking the quasi-long-range order of ultracold quantum Bose gases in two dimensions. If two plates are inserted in the medium, this optical quasi-condensate gives rise to a long-range Casimir-like force, attractive at large distances and repulsive at short distances.  
\end{abstract}
\maketitle

\section{Introduction} 

In its original version, the Casimir force stems from the confinement of the quantum fluctuations of the electromagnetic field: two objects placed in vacuum modify the electromagnetic ground-state energy, which in turn induces an attractive interaction between them \cite{Casimir48, Klimchitskaya09}. 
Beyond this traditional scenario, it was quickly realized that fluctuation-driven forces may arise whenever objects are immersed in a fluctuating environment, which may or may not be made of photons \cite{Kardar99, Gambassi09, Bordag09}. 
Such forces were investigated, e.g., in the vicinity of a critical point in binary liquid mixtures \cite{Garcia02, Fukuto05, Ganshin06, Hertlein08} (critical Casimir effect). In the context of quantum gases, fluctuation-driven forces were also considered for impurities embedded in interacting quantum gases of massive particles 
\cite{Roberts05, Klein05, Recati05, Wachter07, Carusotto17, Dehkharghani18, Pavlov18, Schecter14, Reichert19}. 
This problem is especially interesting in low dimensions, where interacting Bose gases spontaneously form quasi-condensates, whose quantum fluctuations exhibit long-range correlations \cite{Hadzibabic06, Clade09, Petrov00, Mora03}. 
Since, in a Casimir-like scenario, the range of these correlations controls the range of the force, quasi-condensates constitute excellent candidates for the generation of a sizeable interaction between objects. 
In this context, special attention was paid to one-dimensional Bose gases at equilibrium \cite{Schecter14, Reichert19}, where algebraic correlations give rise to long-range Casimir-like forces.

While the notion of condensation seems, at first sight, restricted to massive ultracold gases, many theoretical and experimental efforts have been recently undertaken to describe and observe Bose condensation of light. After seminal observations in polariton systems \cite{Deng02, Kasprzak06},
room-temperature condensates of light were achieved in dye-filled optical microcavities where the confined photons acquire an effective mass and thermalize via their interactions with the dye molecules \cite{Klaers10}. 
Another strategy to thermalize massive photons consists in letting an optical beam propagate in a cavityless, nonlinear Kerr medium. In the paraxial approximation, the propagation is governed by a nonlinear Schr\"odinger equation where the optical axis plays the role of time and the nonlinearity the role of photon interactions \cite{Agrawal95, Rosanov02}. The beam thus behaves as a fluid of light \cite{Carusotto13}, which may thermalize  at long enough propagation time 
\cite{Connaughton05, Sun12, Santic18}.
Another fundamental interest of this setup lies in its two-dimensional nature. This implies that, if condensation cannot exist without cavity, quasi-condensation is on the other hand possible. To our knowledge however, quasi-condensates of light have not yet been considered nor observed experimentally.
Regarding the interaction between fluids of light and matter, the drag forces experienced by an obstacle have been investigated theoretically \cite{Larre15}, and  a recent experiment showed evidence for the suppression of such forces in a photorefractive material \cite{Michel17}. This phenomenon was interpreted as the onset of superfluidity, a concept validated via measurements of the Bogoliubov dispersion of photons in atomic vapors \cite{Vocke15, Fontaine18}. Casimir-like forces in Kerr media, have, on the other hand, little been addressed so far.
\begin{figure}
\centering
\includegraphics[width=0.65\linewidth]{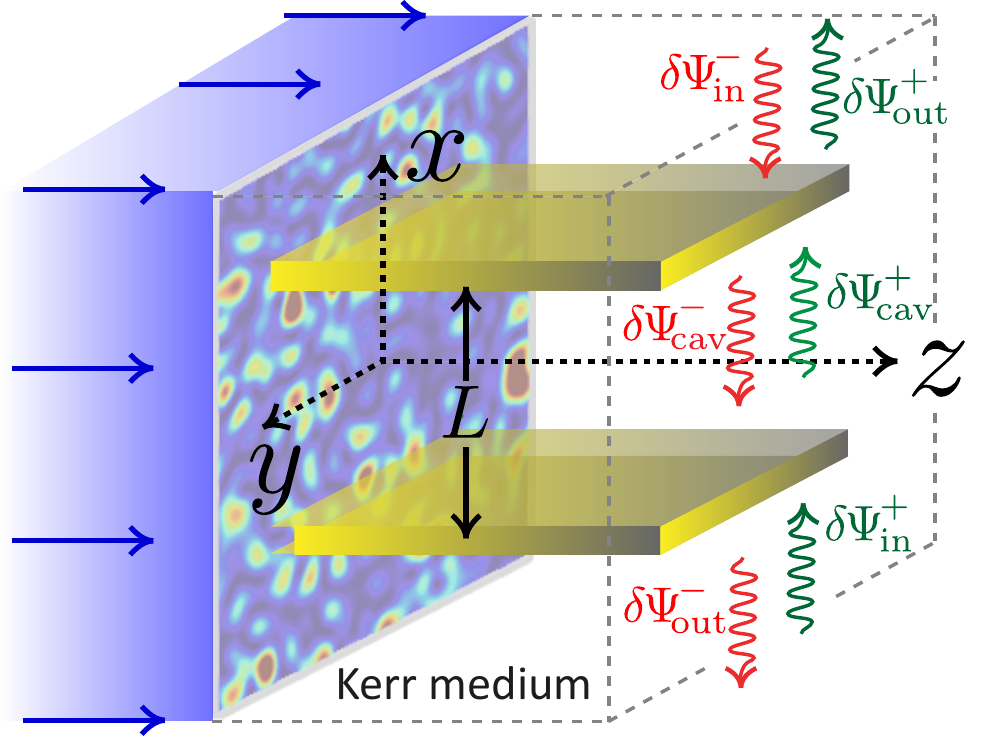}
\caption{
\label{Casimir_scheme}
Upon propagating in a three-dimensional Kerr material, an optical beam with initially small spatial fluctuations develops a transverse algebraic coherence that
induces a long-range Casimir-like pressure between two nearby objects embedded in the medium (here two plates). We describe this pressure within a scattering approach where fluctuations are unitarily reflected by and transmitted through the plates.
}
\end{figure}

In this article, we theoretically show that weakly incoherent optical beams propagating in a Kerr medium over a short distance exhibit a universal, algebraic coherence, mimicking ultracold Bose quasi-condensates in two dimensions. If two objects are immersed in the medium, this long-range coherence leads to an enhanced, long-range Casimir-like force between them.
Our analysis is based on a natural extension of an experimental setup recently used to measure drag forces on dielectric obstacles \cite{Michel17}, and illustrated in Fig. \ref{Casimir_scheme}. A monochromatic light beam carrying  weak transverse spatial fluctuations is let propagate in a Kerr medium in which two plates, parallel to the optical axis $z$, are embedded. Due to the photon interactions pertained to the nonlinear medium, the initial small fluctuations get amplified and, after a short propagation distance, the beam reaches a quasi-stationary prethermal state \cite{Berges04, Langen16, Gring12, Trotzky12, Langen15}. For low enough initial fluctuations, we find that this state exhibits long-range correlations in the transverse plane $(x,y)$, triggering an unconventional Casimir-like pressure which decays algebraically with the plate separation $L$.

\section{Quasi-condensate of light}
Before addressing the complete problem in Fig. \ref{Casimir_scheme}, let us forget the plates for a moment and consider a monochromatic, plane-wave optical beam impinging on a homogeneous, semi-infinite Kerr material at $z=0$. We write the electric field at any point  $(\br_\perp,z)\equiv(x,y,z)$ as $\textbf{E}(\br,z,t)=\mathcal{R} [\Psi(\br_\perp,z)e^{i k_0 z-i\omega t}]\textbf{e}_y$,
where $\omega$ is the  carrier frequency, $k_0=\omega/c$, and $\textbf{e}_y$ is a unit polarization vector along the $y$ axis. In the paraxial approximation, the complex field envelope  $\Psi(\br,z)$ obeys the two-dimensional nonlinear Schr\"odinger equation \cite{Agrawal95, Rosanov02}
\begin{equation}
i\partial_z\Psi(\br_\perp,z)=\left[-\frac{1}{2k_0}\boldsymbol{\nabla}_\perp^2+g|\Psi(\br_\perp,z)|^2\right]
\Psi(\br_\perp,z),
\label{eq:nonlinear_Schro}
\end{equation}
where $g$ controls the strength of the Kerr nonlinearity, assumed to be defocusing, $g>0$.
Suppose now that the incident beam is prepared as a superposition of a uniform background of intensity $I_0$ and a spatially fluctuating speckle field $\phi(\br_\perp)$, $\Psi(\br_\perp,z=0)=\sqrt{I_0}+\epsilon\,\phi(\br_\perp)$.
We describe the latter as a complex, Gaussian random function of two-point correlation $\langle \phi(\br_\perp)\phi^*(\br_\perp+\Delta\br)\rangle=I_0 \gamma(\Delta\br)$, where the brackets refer to statistical averaging. 
For definiteness, in the following we consider a Gaussian correlation, $\gamma(\Delta\br)\equiv \exp(-\Delta\br^2/4\sigma^2)$, with correlation length $\sigma$ \cite{footnote1}. Our main results are, however, independent of this specific choice. From now on, we also mainly focus on the limit $\epsilon\ll 1$ of a weakly incoherent field. This, indeed, 
corresponds to the most interesting configuration where the incident beam mimics a noninteracting, low-temperature Bose gas
undergoing an interaction quench upon entering the nonlinear medium. 
 
The coherence properties of the beam in the material are encoded in the coherence function $g_1(\Delta \br, z)\equiv\langle\Psi(\br_\perp,z)\Psi^*(\br_\perp+\Delta\br,z)\rangle$ \cite{footnote1}.
We have first calculated $g_1$ by numerically propagating the initial state $\Psi(\br_\perp,z=0)$ with Eq. (\ref{eq:nonlinear_Schro}) using a split-step method. 
For the simulations we choose a nonlinearity such that the ratio $\xi/\sigma$ of the healing length $\xi\equiv1/\sqrt{4gI_0 k_0}$ to the speckle correlation length is small, a  condition  required for the nonlinearity to have a significant impact on the beam evolution.
The results  are shown in Fig. \ref{g1_BG_final} against $\Delta r/\sigma$ (dots) for increasing values of $z/z_\text{NL}$, where $z_\text{NL}\equiv1/2gI_0$ is the nonlinear length. 
At $z=0$ (upper black dots),  the coherence function $g_1(\Delta \br, z=0)=I_0\left[1+\epsilon^2\gamma(\Delta\br)\right]$ describes the initial superposition of the  plane wave and small speckle component (solid black curve). 
This structure changes dramatically when $z\ne0$. After a fast, transient evolution over a few tens of $z_\text{NL}$, the overall coherence drops but the short-range component $I_0\epsilon^2\gamma(\Delta\br)$ 
is converted into a long-range, algebraic correlation, as shown in the inset Fig. \ref{g1_BG_final}.
Once this regime has been reached,  $g_1$ also varies rather weakly with $z$ over a spatial range set by the Lieb-Robinson bound $\Delta r= 2c_s z$, where $c_s\equiv\sqrt{gI_0/k_0}$ is the speed of sound.
This phenomenon, known as \textit{prethermalization}, describes a quasi-stationary regime where the beam behaves as a quasi-thermalized, weakly interacting fluid \cite{Berges04, Gring12, Trotzky12, Langen15, Langen16}. 
Since $\epsilon\ll1$, the effective temperature of this state is typically low and the fluid is similar to a quasi-condensate, mimicking the well-known quasi-long-range order of ultracold quantum Bose gases in two dimensions \cite{Hadzibabic06, Clade09, Petrov00, Mora03}. 
Out of the ``light cone'', i.e. for $\Delta r> 2c_s z$, long-range correlations have not yet the time to establish and $g_1$ reaches a plateau reminiscent of the coherent component of the initial beam.
\begin{figure}
\centering
\includegraphics[width=0.95\linewidth]{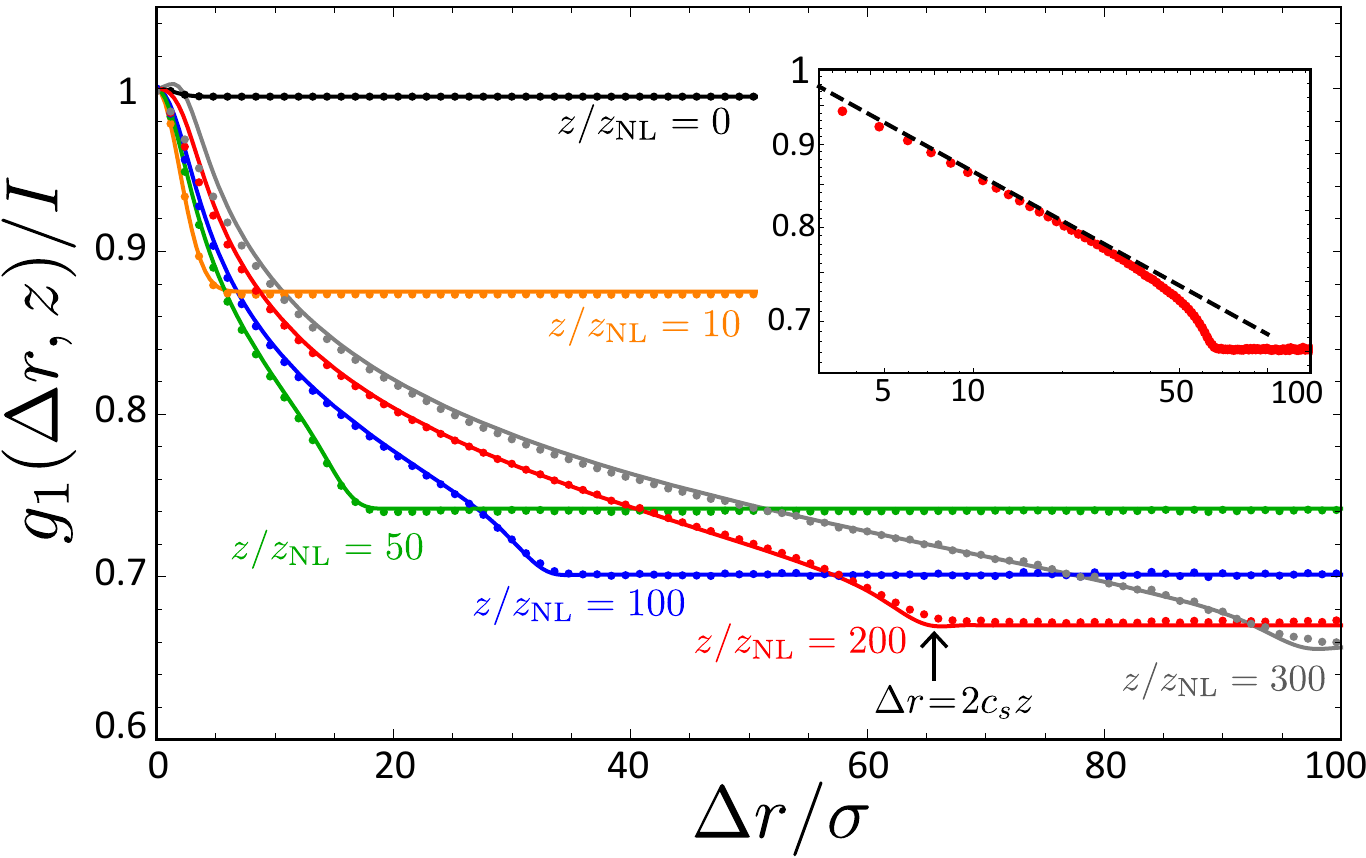}
\caption{
\label{g1function_BG}
Coherence function $g_1$ versus $\Delta r/\sigma$, for increasing values of $z/z_\text{NL}$ and fixed $\epsilon=0.07$ and $\xi/\sigma=0.158$, where $\xi$ is the healing length. Dots are obtained from the numerical resolution of Eq. (\ref{eq:nonlinear_Schro}) with the initial state $\Psi(\br_\perp,z=0)=\sqrt{I_0}+\epsilon\,\phi(\br_\perp)$. Solid curves are Eq. (\ref{g1_BG_final}), including renormalization due to beyond-Bogoliubov corrections. Inset: $g_1$ at $z/z_\text{NL}=200$ in double log scale. The dashed line, Eq. (\ref{Asymptotics_algebraics}), emphasizes the quasi-long-range order within the light cone. 
}
\end{figure}

Theoretically, this behavior is well captured by a time-dependent Bogoliubov description. 
This approach has been previously used to describe the out-of-equilibrium dynamics of quenched, weakly interacting quantum gases \cite{Larre16, Martone18}. Here we adapt it to a classical light beam evolving from the initial state $\Psi(\br_\perp,z=0)$ onwards. Since the beam propagates in an effective two-dimensional space -- $z$ playing the role of a propagation time, see Eq. (\ref{eq:nonlinear_Schro}) --, its phase fluctuations are large. This requires to make use of a density-phase formalism \cite{Mora03},  as detailed in Appendix \ref{appendixA}. The result for $g_1$ is:
\begin{align}
g_1&(\Delta\br,z)=I
\exp\bigg\{-\epsilon^2\int\!\frac{d^2\bq}{(2\pi)^2}
(1-\cos\bq\cdot\Delta\br)\gamma(\bq)\nonumber\\
&\times\bigg[1+\frac{(2gI_0)^2}{2k^2(\bq)}\sin^2k(\bq)z\bigg]\bigg\},
\label{g1_BG_final}
\end{align} 
where $k(\bq)=\sqrt{\bq^2/2k_0[\bq^2/2k_0+2gI_0]}$ is the Bogoliubov dispersion relation, $\gamma(\bq)=\int d^2\br_\perp \gamma(\Delta\br)e^{-i\bq\cdot\Delta\br}$ is the speckle power spectrum, and $I=I_0(1+\epsilon^2)=\langle|\Psi(\br_\perp,z)|^2\rangle$ is the total light intensity, which is conserved during the evolution. 
Note that, at $z=0$, Eq. (\ref{g1_BG_final}) well reduces  to $I_0\left[1+\epsilon^2\gamma(\Delta\br)\right]$ since $\epsilon\ll1$. 
While the Bogoliubov approach is generally valid at small $z$, as $z$ increases interactions between quasiparticles become relevant and should be accounted for \cite{Regemortel18}. In the prethermal regime we are interested in however, their effect at $\Delta r\gg\sigma$ is very well captured by a simple renormalization of $I$, as explained in Appendix \ref{sec:fit}. Using this procedure,  the agreement between Eq. (\ref{g1function_BG}) and the numerical data is excellent over two orders of magnitude of $z/z_\text{NL}$, as seen in Fig. \ref{g1function_BG}.
The algebraic decay of $g_1$, visible when $z\gg z_\text{NL}$, is a consequence of the large phase fluctuations of the beam in the nonlinear medium, stemming from the infrared divergence $1/k^2(\bq)\sim \bq^{-2}$ in Eq. (\ref{g1_BG_final}). They yield the asymptotic law:
\begin{align}
g_1(\Delta\br,z\gg z_\text{NL})\simeq I 
\left(\frac{\sigma}{\Delta r}\right)^\alpha,
\label{Asymptotics_algebraics}
\end{align}
with an exponent $\alpha=\epsilon^2\sigma^2/2\xi^2$. The algebraic law (\ref{Asymptotics_algebraics}) is shown in the inset of Fig. \ref{g1function_BG} (dashed curve). It signals the formation of a quasi-condensate of light. 
Eq. (\ref{Asymptotics_algebraics}) holds up to the light-cone  bound $\Delta r= 2c_s z$. Out of the light cone, the coherence function saturates at $g_1\sim I(\sigma/c_s z)^\alpha$.
We stress that the emergence of long-range coherence discussed here does not rely on non-local effects \cite{Fusaro17}. It spontaneously emerges even with local interactions, provided one starts from a \emph{weakly} incoherent beam.

\section{Casimir-like force}

The long-range coherence exhibited by optical beams in the prethermal regime makes the configuration of Fig. \ref{Casimir_scheme} promising for realizing a sizeable Casimir-like force. To confirm this intuition, we now
add the plates and explore the fluctuations-induced interaction between them. To calculate this interaction, we make use of a scattering approach to  Casimir forces, in which the effect of the plates is described in terms of the transmission and reflection of field fluctuations in the absence of coupling, assuming unitarity only \cite{Jaekel91} (in the configuration of Fig. \ref{Casimir_scheme}, the uniform mean-field component $\langle\Psi(z)\rangle$ does not yield any force). The first step of this approach consists in decomposing the field fluctuations in the three regions delineated by the plates into components moving forward and backward along the $x$ axis, as shown in Fig. \ref{Casimir_scheme}. 
In the two outer regions, we express the incoming field fluctuations as $\delta \Psi_\text{in}^{\pm}(\br_\perp,z)=\int_{q_x\small\gtrless0} d^2\bq/(2\pi)^2 \delta\Psi(\bq,z)e^{i\bq\cdot\br_\perp}$, where the Fourier components $\delta\Psi(\bq,z)\equiv\Psi(\bq,z)-\langle\Psi(\bq,z)\rangle$ refer to the beam fluctuations  in the absence of plates. The scattered fields then follow from $(\delta\Psi_\text{out}^+,\delta\Psi_\text{out}^-)=S(\delta\Psi_\text{in}^+,\delta\Psi_\text{in}^-)$ and $(\delta\Psi_\text{cav}^+,\delta\Psi_\text{cav}^-)=R(\delta\Psi_\text{in}^+,\delta\Psi_\text{in}^-)$ where $S$ and $R$ are, respectively, the scattering and resonance matrices of the cavity formed by the plates. The explicit expression of $S$ and $R$ is given in Appendix \ref{AppendixB}. 
With the fields in the three regions expressed in terms of the components of $S$ and $R$, we then evaluate  the average radiation pressures on each side of a given plate, the difference of which defining the Casimir pressure. The radiation pressure is given by the diagonal component $T_{xx}$ of the stress tensor of the fluid of light \cite{Pavloff02}. Within the Bogoliubov approximation and using the unitary transformation $\Psi\to \Psi\exp(igI_0z)$, we find for instance  (see Appendix \ref{AppendixB}) that
the energy flux associated with the incoming field $\delta\Psi_\text{in}^{+}(\br_\perp,z)$ exerts a pressure
\begin{align}
&T_{xx}(\delta\Psi_\text{in}^+)=\frac{\epsilon_0}{2k_0}\int_{q_x\!>0}\!\!\frac{d^2\bq}{(2\pi)^2}
\left[\frac{q_x^2}{2k_0}\langle|\delta\Psi(\bq,z)|^2\rangle+\right.\nonumber\\
&
\left.\Im\langle\delta\Psi(\bq,z)\partial_z\delta\Psi^*(\bq,z)\rangle
\right],
\label{Txx_psiinp}
\end{align}
where $\epsilon_0$ if the vacuum permittivity  and $\bq=(q_x,q_y)$.
\begin{figure}
\centering
\includegraphics[width=0.95\linewidth]{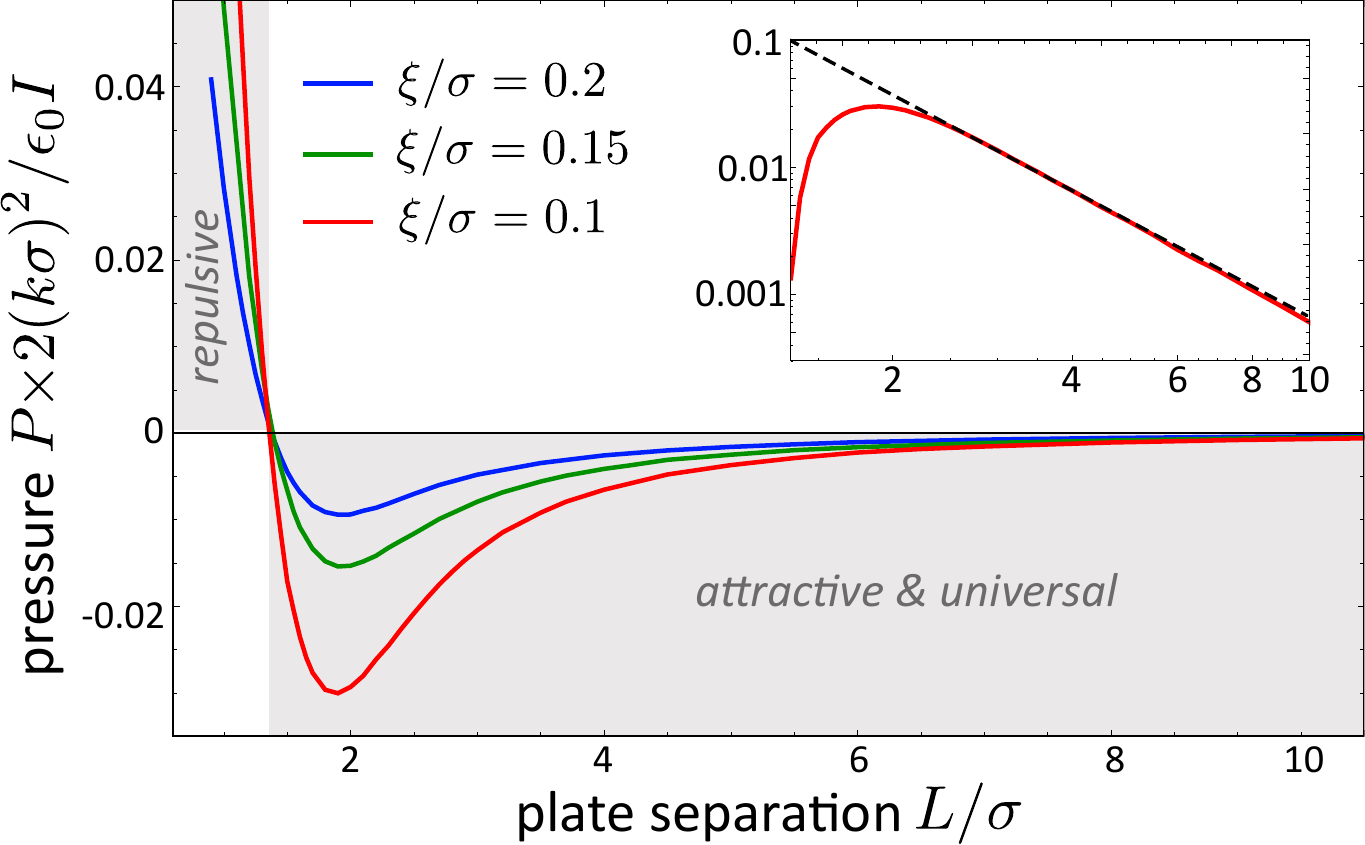}
\caption{
\label{Casimir_force}
Casimir pressure, Eq. (\ref{P_Casimir_final}), as a function of the plate separation, for $\epsilon=0.1$. The pressure is attractive at large separation, where it exhibits an algebraic decay associated with the long-range coherence of the prethermal fluid of light, and repulsive at short separation. Its  magnitude increases as the ratio of the healing length $\xi$ to the speckle correlation length $\sigma$ decreases, i.e. as the nonlinearity gets stronger. Inset: pressure for $\xi/\sigma=0.1$ in double log scale, emphasizing the algebraic decay. The dashed line is Eq. (\ref{P_asymptotics}).
}
\end{figure}
This radiation pressure has two contributions. The first is given by the transverse energy $q_x^2/2k_0$ of the paraxial photons, weighted by the spectrum  $\langle|\delta\Psi(\bq,z)|^2\rangle$ of their fluctuations. The second, $\Im\langle\delta\Psi(\bq,z)\partial_z\delta\Psi^*(\bq,z)\rangle$, is the current fluctuation spectrum. It stems from the non-equilibrium nature of the evolution and is usually absent  in equilibrium configurations \cite{Pavloff02}. 
By relating the reflected fluctuations $\delta\Psi_\text{out}^-$ to $\delta\Psi_\text{in}^+$ and $\delta\Psi_\text{in}^-$ using the scattering matrix $S$ and invoking unitarity, we then find $T_{xx}(\delta\Psi_\text{out}^-)=T_{xx}(\delta\Psi_\text{in}^+)$. Calculation of the radiation pressure inside the cavity, finally, follows the same lines but now involves the elements of the resonance matrix $R$. The Casimir pressure $P=T_{xx}(\delta\Psi_\text{cav}^+)+T_{xx}(\delta\Psi_\text{cav}^-)-T_{xx}(\delta\Psi_\text{in}^+)-T_{xx}(\delta\Psi_\text{out}^-)$ then reads
\begin{align}
&P=\frac{2\epsilon_0}{k_0}\Re\int_{q_x\!>0}\!\!\frac{d^2\bq}{(2\pi)^2}
\frac{r^2(\bq)e^{2iq_xL}}{1-r^2(\bq)e^{2iq_xL}}\nonumber\\
&\times\left[\frac{q_x^2}{2k_0}\langle|\delta\Psi(\bq,z)|^2\rangle+
\Im\langle\delta\Psi(\bq,z)\partial_z\delta\Psi^*(\bq,z)\rangle
\right],
\label{P_Casimir_general}
\end{align}
where $r(\bq)$ denotes the reflection coefficient of a single plate
in the direction $\bq$, and $L$ is the plate separation. The Casimir pressure thus naturally appears as the noise (density plus current) spectrum of the fluid of light, weighted by the admittance of the cavity
, summed over all possible scattering directions $\bq$. 
Eq. (\ref{P_Casimir_general}) can be further simplified by noting that, in the paraxial approximation, the fluctuations are essentially scattered at grazing incidence. It follows that  $r^2(\bq)\simeq 1$ whatever the nature of the material the plates are made of. Eq. (\ref{P_Casimir_general}) can then be reformulated in position space as
\begin{align}
P=-\frac{\epsilon_0}{2}\sum_{n=0}^\infty\left[\frac{\partial^2g_1(\Delta\br,z)}{k_0^2\partial{\Delta x}^2}\!+\!g_1^j(\Delta\br,z)\right]
_{\!\begin{smallmatrix*}[l]
 \Delta x=2L(n+1)\\
  \Delta y=0
\end{smallmatrix*}}
\label{P_Casimir_final}
\end{align}
where the sum runs over all resonance spatial frequencies of the cavity.
In this relation, the first term in the right-hand side involves the coherence function (\ref{g1_BG_final}). This term dominates at large separation $L\gg\sigma$, where the phase fluctuations of the beam make $g_1$ long-range.
We show in Appendix \ref{sec:current_spectrum} that the current correlator $g_1^j(\Delta\br,z)\equiv2\Im\langle\delta\Psi^*(\br_\perp,z)\partial_z\delta\Psi(\br_\perp+\Delta\br,z)\rangle/k_0$, on the other hand, is essentially governed by the intensity fluctuations of the beam, which are typically small when $\epsilon\ll 1$. Its contribution is thus  important at short scale $L\lesssim\sigma$ only.

The Casimir pressure (\ref{P_Casimir_final}), calculated with the Bogoliubov theory, is shown  in Fig. \ref{Casimir_force} as a function of the plate separation $L$, in the prethermal regime $z\gg z_\text{NL}$ where it is essentially independent of $z$.
Its most remarkable feature is the behavior at large separation. The latter is governed by the first term in the right-hand side of Eq. (\ref{P_Casimir_final}), with $g_1$ given by Eq. (\ref{Asymptotics_algebraics}) and evaluated at the lowest resonance, $n=0$:
\begin{align}
P(L\gg\sigma)\sim-\dfrac{\epsilon_0 I_0}{(k_0\sigma)^2}\alpha(\alpha+1)\left(\frac{\sigma}{L}\right)^{\alpha+2}.
\label{P_asymptotics}
\end{align}
Eq. (\ref{P_asymptotics}) is a central result of the article. It shows that at large separation the pressure is attractive and decays algebraically. The asymptotic law (\ref{P_asymptotics}) is compared with the exact formula (\ref{P_Casimir_final}) in the inset of Fig. \ref{Casimir_force}.
The algebraic decay is governed by 
the exponent $\alpha=\epsilon^2\sigma^2/2\xi^2$, which can be either larger or smaller than 1 since both $\epsilon\ll1$ and $\xi/\sigma\ll1$. 
In particular, when $\alpha>1$, the pressure is \textit{much larger} than the pressure $P\sim -\epsilon_0 I_0/(k_0\sigma)^2\gamma^{\prime\prime}(2L)$ that would result from the use of a fully developed speckle, i.e. $\Psi(\br_\perp,z)=\phi(\br_\perp)$. Furthermore, the decay (\ref{P_asymptotics}) is \textit{universal},  in the sense that it only depends on the small set of parameters $(\xi,\sigma,\epsilon)$, but not on the specific shape of $\gamma(\Delta\br)$. 
At small separation $L\lesssim \sigma$, the pressure (\ref{P_Casimir_final}) becomes governed by the current correlator and turns repulsive, as seen in Fig. \ref{Casimir_force}. Its $L$ dependence at such short scale is nonuniversal in general, i.e. it depends on the  shape of $\gamma(\Delta\br)$ (see Appendix \ref{sec:current_spectrum}).
Fig. \ref{Casimir_force} and Eq. (\ref{P_asymptotics}) also reveal that the overall magnitude of the pressure increases with decreasing $\xi/\sigma$.
This result can be understood as follows. When $\xi\ll\sigma$, the speckle spectrum selects only the low (phonon-like) Bogoliubov modes $|\bq|\xi\ll1$, responsible for the algebraic decay of the coherence function and a sizable Casimir force. In contrast, when $\xi/\sigma\gtrsim1$ the speckle spectrum also captures particle-like modes $|\bq|\xi\gtrsim1$. Since these modes describe purely non-interacting particles, their coherence function hardly evolves from its form at $z=0$, which carries  small fluctuations and therefore leads to a small force.

\begin{figure}[h]
\centering
\includegraphics[width=0.9\linewidth]{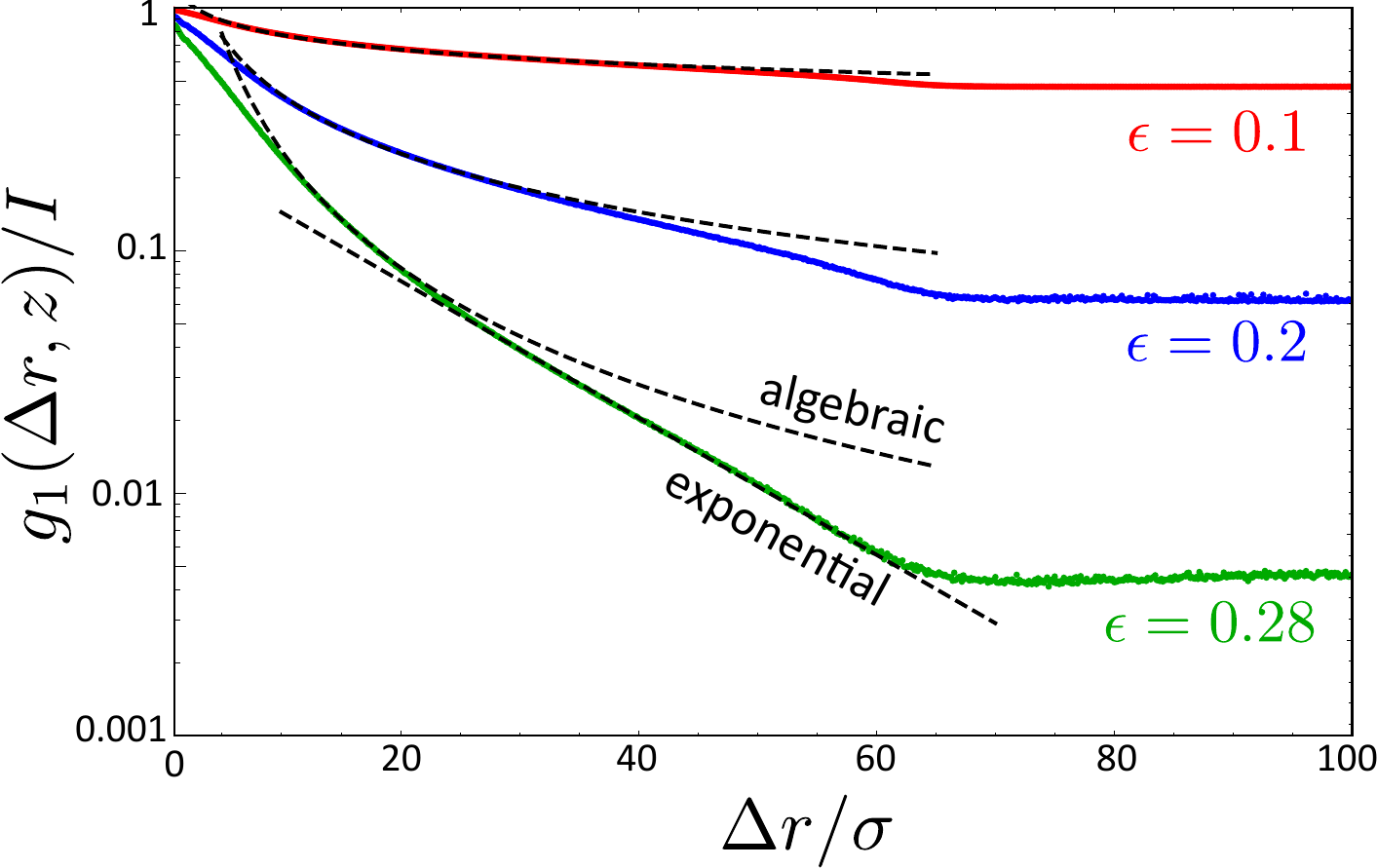}
\caption{
\label{g1function_larger_eps}
Coherence function versus $\Delta r/\sigma$ for increasing values of $\epsilon$ and  fixed $\xi/\sigma=0.158$ and $z/z_\text{NL}=200$, obtained from the numerical resolution of Eq. (\ref{eq:nonlinear_Schro}). While the algebraic law (\ref{Asymptotics_algebraics}) is observed 
at small $\epsilon$ (dashed curves), a crossover to an exponential decay (lower dashed line) shows up at larger $\epsilon$.
}
\end{figure}

We finally comment on the role of the parameter $\epsilon$, which controls the amount of fluctuations in the incident beam. At small $\epsilon$, the effective temperature of the prethermal regime is small, so that the fluid of light effectively behaves as a low-temperature interacting Bose gas in two dimensions, i.e., a quasi-condensate.
By analogy, a larger $\epsilon$ will  describe a gas of temperature typically above the quasi-condensation threshold, i.e., of exponentially small coherence \cite{Kosterlitz73}.
This qualitative picture is confirmed by numerical simulations of $g_1$ shown in Fig. \ref{g1function_larger_eps}. As $\epsilon$ increases, the algebraic behavior of $g_1$ turns to an exponential decay, making the pressure (\ref{P_Casimir_final}) much weaker.
In other words, and perhaps counterintuitively, as far as the Casimir pressure is concerned it is much more interesting  to inject small fluctuations and let the Kerr medium amplify them than using strong fluctuations for the start.

\section{Conclusion}

Let us conclude with experimental considerations. In atomic vapors illuminated slightly away from resonance, nonlinearities such that $z_\text{NL}\simeq 1$ mm  and $\xi\simeq 10\, \mu$m can be reached \cite{Santic18}. For a cell length $z=7$ cm, this corresponds to $z/z_\text{NL}\simeq 70$ and $2c_s z\simeq 1.4$ mm for the Lieb-Robinson bound, much larger than the speckle correlation $\sigma$, usually on the order of a few tens of microns. A large window $2c_s z/\sigma $ of two or three orders of magnitude can thus be realized, making the long-range behavior of $g_1$ observable under rather reasonable conditions. Measuring the Casimir force is more difficult as it requires to distinguish it from stray forces unrelated to fluctuations. One example 
are the field perturbations induced by the plate edges at $z=0$, which may create an additional deterministic force. This contribution could, however, be 
removed by prior measurement of the force created by the non-fluctuating background, or reduced by using smooth plate profiles near the interface.

\section*{Acknowledgments}

NC thanks Robin Kaiser, Dominique Delande, Romain Gu\'erout, Serge Reynaud, Pierre-\'Elie Larr\'e and Tom Bienaim\'e for useful discussions. NC and SP acknowledge the Agence Nationale de la Recherche (grants ANR-19-CE30-0028-01 CONFOCAL and ANR-16-ACHN-0027 C-FLigHT) for financial support.

\appendix
\section{Bogoliubov theory of out-of-equilibrium optical beams}
\label{appendixA}

To derive the coherence function (\ref{g1_BG_final}), we use a ``density-phase'' formalism \cite{Mora03}: we write the complex field as $\Psi(\br_\perp,z)=\sqrt{I+\delta I(\br_\perp,z)}\exp[-i g I_0 z+i\theta(\br_\perp,z)]$, insert this Ansatz into Eq. (\ref{eq:nonlinear_Schro}) and linearize with respect to the intensity fluctuations $\delta I(\br_\perp,z)$. This yields
\begin{align}
&\partial_z \delta I(\br_\perp,z)=-\boldsymbol{\nabla}^2\theta(\br_\perp,z)/k_0\label{eq:BGdGennes1}\\
&\partial_z  \theta(\br_\perp,z)=\boldsymbol{\nabla}^2 \delta I(\br_\perp,z)/4k_0I_0-g\delta I(\br_\perp,z).
\label{eq:BGdGennes2}
\end{align} 
At this stage, let us mention that the linearization procedure is permitted because the intensity fluctuations of the initial state $\Psi(\br_\perp,z=0)=\sqrt{I_0}+\epsilon\,\phi(\br_\perp)$ are indeed small when $\epsilon\ll1$, see Eq. (\ref{deltaI0}) below. Note also that when writing Eqs. (\ref{eq:BGdGennes1}) and (\ref{eq:BGdGennes2}) we did \textit{not} linearize with respect to the phase $\theta$, whose fluctuations are typically large in low-dimensional interacting systems \cite{Mora03}.
To find $\delta I$ and $\theta$, we diagonalize Eqs. (\ref{eq:BGdGennes1}) and (\ref{eq:BGdGennes2}) in Fourier space, using the Fourier transform $\delta I(\bq,z)\equiv \int d^2\br \delta I(\br_\perp,z)e^{-i\bq\cdot\br_\perp}$ and similarly for $\theta(\bq,z)$, and solve the resulting differential equations with the initial conditions
\begin{align}
\delta I(\br_\perp,z=0)=2\epsilon\sqrt{I_0}\,\phi_r(\br_\perp)\label{deltaI0}\\
 \theta(\br_\perp,z=0)=
\epsilon\,\phi_i(\br_\perp)/\sqrt{I_0},
\end{align} 
where $\phi_r(\br_\perp)\equiv\Re\phi(\br_\perp)$ and $\phi_i(\br_\perp)\equiv\Im\phi(\br_\perp)$. This leads to
\begin{align}
&\delta I(\br_\perp,z)=\epsilon\sqrt{I_0}
\int\!\frac{d^2\bq}{(2\pi)^2}\left[\phi_r(\bq)+i\frac{K(\bq)}{k(\bq)}\phi_i(\bq)\right]
\label{BG_expressions1}\nonumber\\
&\times\exp[-ik(\bq)z+i\bq\cdot\br_\perp]+\text{c.c.}\\
&\theta(\br_\perp,z)=
\frac{\epsilon}{2i\sqrt{I_0}}
\int\!\frac{d^2\bq}{(2\pi)^2}
\left[\frac{k(\bq)}{K(\bq)}\phi_r(\bq)+i\phi_i(\bq)\right]\nonumber\\
&\times\exp[-ik(\bq)z+i\bq\cdot\br_\perp]+\text{c.c.},
\label{BG_expressions2}
\end{align} 
where $K(\bq)=\bq^2/2k_0$ and $k(\bq)=\sqrt{K(\bq)[K(\bq)+2gI_0]}$ is the Bogoliubov dispersion relation. The Fourier components $\phi_r(\bq)$ and $\phi_i(\bq)$ follow a Gaussian statistics, and their correlators obey $\langle\phi_r(\bq)\phi_r^*(\bq')\rangle=\langle\phi_i(\bq)\phi_i^*(\bq')\rangle=(2\pi)^2\delta(\bq-\bq')\gamma(\bq)/2$ and $\langle\phi_r(\bq)\phi_i^*(\bq')\rangle=0$, with $\gamma(\bq)$ the speckle power spectrum. It follows that
\begin{align}
g_1(\Delta\br,z)&=\ I
\exp\left\{-\frac{1}{2}
\langle[\theta(\br_\perp,z)-\theta(\br_\perp+\Delta\br,z)]^2\rangle\right.\nonumber\\
&\left.-\frac{1}{8}\langle[\delta I(\br_\perp,z)-\delta I(\br_\perp+\Delta\br,z)]^2\rangle
\right\},
\label{g1_thetaI}
\end{align}
which is the same expression as in quantum gases \cite{Mora03}. We finally obtain Eq. (\ref{g1_BG_final}) by inserting Eqs. (\ref{BG_expressions1}) and (\ref{BG_expressions2}) into Eq. (\ref{g1_thetaI}).

\section{Fit of numerical coherence functions}
\label{sec:fit}

While Eq. (\ref{g1_BG_final}) faithfully models the coherence function at small $z$, higher-order nonlinear contributions arise as the beam propagates deeper in the Kerr medium. Physically, these corrections describe interactions between the Bogoliubov quasiparticles and become increasingly important as $z/z_\text{NL}$ increases. Eventually, they entail the full thermalization of the fluid of light \cite{Regemortel18}. While their complete description at any $z_\text{NL}$ is a formidable task, in the present work we are interested in the prethermal regime where quasiparticle interactions only bring small corrections to Eq. (\ref{g1_BG_final}). We model them by adding a phenomenological parameter $\beta$ according to 
\begin{align}
g_1&(\Delta\br,z)\simeq I
\exp\bigg\{-\epsilon^2\int\!\frac{d^2\bq}{(2\pi)^2}
(1-\cos\bq\cdot\Delta\br)\gamma(\bq)\nonumber\\
&\times\bigg[1+\frac{(2gI_0)^2}{2k^2(\bq)}\sin^2k(\bq)z
+\beta(z)\bigg]\bigg\}.
\label{g1_BG_final_beta}
\end{align} 
In this prescription, which preserves normalization, the three terms $1$, $(2gI_0)^2\sin^2k(\bq)z/2k^2(\bq)$ and $\beta(z)$ within the exponential can be interpreted as the zeroth-, first- and second-order contributions of a perturbation expansion of $\ln g_1/I$. Guided by the universality of the algebraic decay of $g_1$ in the prethermal regime, captured by the Bogoliubov contribution, we assume $\beta$ to depend on $z$ only. This turns out to be an excellent approximation, as seen in Fig. \ref{g1function_BG} where the solid curves are obtained by fitting Eq. (\ref{g1_BG_final_beta}) to the numerical data, with $\beta(z)$ as the only fit parameter. 
In practice, we find that the prescription (\ref{g1_BG_final_beta}) typically works up to $z/z_\text{NL}\sim 1000$. At larger $z$, the system starts deviating much from the prethermal regime and a more general kinetic theory is needed \cite{Regemortel18}.

\section{Derivation of the Casimir pressure}
\label{Sec:Casimirforce}
\label{AppendixB}

To derive the Casimir pressure exerted by the fluctuations of the fluid of light on the plates, we first express the field fluctuations in the three regions delineated by the plates (see Fig. \ref{Casimir_scheme}). The incoming fields, first, read
\begin{align}
\delta \Psi_\text{in}^{+}(\br_\perp,z)=\!\int_{q_x>0}\! \frac{d^2\bq}{(2\pi)^2} \delta\Psi^+(\bq,z)e^{iq_x x+ i q_y y}
\label{Psi_in_plus}
\end{align}
and
\begin{align}
\delta \Psi_\text{in}^{-}(\br_\perp,z)=\!\int_{q_x>0}\! \frac{d^2\bq}{(2\pi)^2}  \delta\Psi^-(\bq,z)e^{-iq_x x+ i q_y y},
\label{Psi_in_minus}
\end{align}
where $\delta\Psi^+(\bq,z)\equiv\Psi(\bq,z)-\langle\Psi(\bq,z)\rangle$ (with $\bq=(q_x,q_y)$) and $\delta\Psi^-(\bq,z)\equiv\Psi(-q_x,q_y,z)-\langle\Psi(\bq, z)\rangle$ are the beam fluctuations without the plates, i.e. with intensity and phase fluctuations given by Eqs. (\ref{BG_expressions1}) and (\ref{BG_expressions2}). 
The field fluctuations scattered by the plates follow from $(\delta\Psi_\text{out}^+,\delta\Psi_\text{out}^-)=S(\delta\Psi_\text{in}^+,\delta\Psi_\text{in}^-)$ and $(\delta\Psi_\text{cav}^+,\delta\Psi_\text{cav}^-)=R(\delta\Psi_\text{in}^+,\delta\Psi_\text{in}^-)$, where $S$ and $R$ are the scattering and resonance matrices of the cavity. In details:
\begin{align}
\delta \Psi_\text{out}^{+}(\br_\perp,z)=&\!\int_{q_x\small>0}\! \frac{d^2\bq}{(2\pi)^2}
[S_{11}(\bq)\delta\Psi^+(\bq,z)\nonumber\\
&+S_{12}(\bq)\delta\Psi^-(\bq,z)]
e^{iq_x x+iq_y y},
\label{Psi_out_plus}
\end{align}
\begin{align}
\delta \Psi_\text{out}^{-}(\br_\perp,z)=&\!\int_{q_x\small>0}\! \frac{d^2\bq}{(2\pi)^2}
[S_{21}(\bq)\delta\Psi^+(\bq,z)\nonumber\\
&+S_{22}(\bq)\delta\Psi^-(\bq,z)]
e^{-iq_x x+iq_y y},
\label{Psi_out_minus}
\end{align}
\begin{align}
\delta \Psi_\text{cav}^{+}(\br_\perp,z)=&\!\int_{q_x\small>0}\! \frac{d^2\bq}{(2\pi)^2}
[R_{11}(\bq)\delta\Psi^+(\bq,z)\nonumber\\
&+R_{12}(\bq)\delta\Psi^-(\bq,z)]
e^{iq_x x+iq_y y},
\label{Psi_cav_plus}
\end{align}
and
\begin{align}
\delta \Psi_\text{cav}^{-}(\br_\perp,z)=&\!\int_{q_x\small>0}\! \frac{d^2\bq}{(2\pi)^2}
[R_{21}(\bq)\delta\Psi^+(\bq,z)\nonumber\\
&+R_{22}(\bq)\delta\Psi^-(\bq,z)]
e^{-iq_x x+iq_y y}.
\label{Psi_cav_minus}
\end{align}
The coefficients $R_{ij}$ and $S_{ij}$ of the resonance and scattering matrices of the cavity depend on  the reflection and transmission coefficients of a single plate, $r$ and $t$, which obey the  unitarity conditions $|r|^2+|t|^2=1$ and $r t^*+t r^*=0$ (to lighten the notations we omit the $\bq$  dependence of $r$ and $t$) \cite{Jaekel91}:
\begin{align}
S=\frac{1}{d}
\begin{pmatrix} 
t^2 & dre^{-i |q_x| L}\!+\!t^2 re^{i |q_x| L} \\
dre^{-i |q_x| L}\!+\!t^2 re^{i |q_x| L} & t^2
\end{pmatrix}
\end{align}
and
\begin{align}
R=\frac{1}{d}
\begin{pmatrix} 
t & tre^{i |q_x| L} \\
tre^{i |q_x| L} & t
\end{pmatrix},
\label{Rmatrix}
\end{align}
where $d\equiv 1-r^2e^{2i |q_x|L}$.

We are now in position to compute the radiation pressures exerted by the various fluctuation components on the plates. To achieve this goal, we make use of the stress tensor of the fluid of light \cite{Pavloff02}:
\begin{align}
T_{xx}(\Psi)=\frac{\epsilon_0}{2}
\!\left[-\frac{\Im \Psi^*\partial_z\Psi}{k_0}+\frac{|\partial_x \Psi|^2}{2k_0^2}-\frac{g|\Psi|^4}{2k_0}
\right].
\label{Stress_tensor_general}
\end{align}
Note that when $g\to 0$ Eq. (\ref{Stress_tensor_general}) coincides with the paraxial limit of the usual time-averaged Maxwell stress tensor of electromagnetic waves in free space \cite{Landau60}. 
To implement Eq. (\ref{Stress_tensor_general}) within a linear scattering theory, it is first required to quadratize $T_{xx}$. In the density-phase formulation, this can be achieved by  noting that, for small intensity fluctuations, $|\Psi|^4\simeq 2I|\Psi|^2-I^2$.
Inserting then Eq. (\ref{Psi_in_plus}) into the quadratized version of (\ref{Stress_tensor_general}) and averaging over the statistics of the speckle, we find, after redefining $\Psi\to \Psi\exp(igI_0z)$:
\begin{align}
&T_{xx}(\delta\Psi_\text{in}^+)=\frac{\epsilon_0}{2k_0}\int_{q_x\!>0}\!\!\frac{d^2\bq}{(2\pi)^2}
\left[\frac{q_x^2}{2k_0}\langle|\delta\Psi(\bq,z)|^2\rangle+\right.\nonumber\\
&
\left.\Im\langle\delta\Psi(\bq,z)\partial_z\delta\Psi^*(\bq,z)\rangle
\right],
\label{Txx_in_plus}
\end{align}
which is Eq. (\ref{Txx_psiinp}). A similar calculation based on Eq. (\ref{Psi_out_minus}) yields 
\begin{align}
&T_{xx}(\delta\Psi_\text{out}^-)=\frac{\epsilon_0}{2k_0}\int_{q_x\!>0}\!\!\frac{d^2\bq}{(2\pi)^2}
(|S_{21}|^2+|S_{22}|^2)\times\nonumber\\
&
\left[\frac{q_x^2}{2k_0}\langle|\delta\Psi(\bq,z)|^2\rangle+\Im\langle\delta\Psi(\bq,z)\partial_z\delta\Psi^*(\bq,z)\rangle
\right],
\label{Txx_out_minus}
\end{align}
which equals $T_{xx}(\delta\Psi_\text{in}^+)$ by virtue of the unitarity of the scattering matrix: $|S_{21}|^2+|S_{22}|^2=1$. The same calculation with Eqs. (\ref{Psi_cav_plus}) and (\ref{Psi_cav_minus}), finally, gives
\begin{align}
&T_{xx}(\delta\Psi_\text{cav}^+)\!+\!T_{xx}(\delta\Psi_\text{cav}^-)\!=\!
\frac{\epsilon_0}{2k_0}\int_{q_x\!>0}\!\!\frac{d^2\bq}{(2\pi)^2}
\sum_{i,j=1,2}|R_{ij}|^2\nonumber\\
&\times\left[\frac{q_x^2}{2k_0}\langle|\delta\Psi(\bq,z)|^2\rangle\!+\!\Im\langle\delta\Psi(\bq,z)\partial_z\delta\Psi^*(\bq,z)\rangle
\right].
\label{Txx_cav}
\end{align}
The prefactor involving the components of the resonance matrix can be explicitly computed from Eq. (\ref{Rmatrix}):
\begin{equation}
\frac{1}{2}\sum_{i,j}|R_{ij}|^2=1+2\Re\frac{r^2e^{2i |q_x|L}}{1-r^2e^{2i |q_x|L}}.
\label{Admittance_func}
\end{equation}
From Eqs. (\ref{Txx_in_plus}), (\ref{Txx_out_minus}), (\ref{Txx_cav}) and (\ref{Admittance_func}), we finally obtain Eq. (\ref{P_Casimir_final}) for the Casimir pressure $P\equiv T_{xx}(\delta\Psi_\text{cav}^+)+T_{xx}(\delta\Psi_\text{cav}^-)-T_{xx}(\delta\Psi_\text{in}^+)-T_{xx}(\delta\Psi_\text{out}^-)$.

\section{Current fluctuation spectrum}
\label{sec:current_spectrum}

In addition to the coherence function $g_1$, the  Casimir pressure (\ref{P_Casimir_final}) also involves the current correlator $g_1^j(\Delta\br,z)\equiv2\Im\langle\delta\Psi^*(\br_\perp,z)\partial_z\delta\Psi(\br_\perp+\Delta\br,z)\rangle/k_0$, where $\delta\Psi(\br_\perp,z)\equiv \Psi(\br_\perp,z)-\langle\Psi(\br_\perp,z)\rangle$.
To calculate it, we proceed as for $g_1$: we  linearize with respect to intensity fluctuations and use the Gaussian statistics of $\phi_r$ and $\phi_i$. This leads to
\begin{align}
&g_1^j(\Delta\br,z)=\frac{1}{k_0}\exp\Big[-\frac{1}{2}\langle[\theta(\br_\perp,z)-\theta(\br_\perp+\Delta\br,z)]^2\rangle\Big]\nonumber\\
&\times\left\{\langle\partial_z \theta(\br_\perp+\Delta\br)[\delta I(\br_\perp,z)+\delta I(\br_\perp+\Delta\br,z)]\rangle\right.\nonumber\\
&-\left.
\langle\partial_z\delta I(\br_\perp+\Delta\br)[\theta(\br_\perp,z)-\theta(\br_\perp+\Delta\br,z)]\rangle
\right\}\nonumber\\
&-\frac{1}{k_0}\exp\big[
-\langle\theta^2(\br_\perp,z)\rangle\big]
\left\{\langle\partial_z\delta I(\br_\perp+\Delta\br)\theta(\br_\perp+\Delta\br)\rangle\right.\nonumber\\
&\left.+\langle\partial_z \theta(\br_\perp+\Delta\br)\delta I(\br_\perp+\Delta\br,z)\rangle\right\}.
\label{g1j_general}
\end{align}
To leading order, the exponential prefactors are given by $\exp\Big[-\langle[\theta(\br_\perp,z)-\theta(\br_\perp+\Delta\br,z)]^2/2\rangle\Big]\!\simeq\! g_1(\Delta\br,z)/I$, and 
\begin{align}
&\exp\left[-\langle\theta^2(\br_\perp,z)\rangle\right]
=g_1(\Delta\br\to\infty,z)/I\\
&=
\exp\bigg\{-\epsilon^2\int\!\frac{d^2\bq}{(2\pi)^2}\gamma(\bq)
\bigg[1+\frac{(2gI_0)^2}{2k^2(\bq)}\sin^2k(\bq)z\bigg]\bigg\}.\nonumber
\end{align}
The intensity-phase correlators, finally, are evaluated from Eqs. (\ref{BG_expressions1}) and (\ref{BG_expressions2}). The final expression of $g_1^j$ is:
\begin{align}
g_1^j(\Delta\br,z)=&g_1(\Delta\br,z)[F(\Delta\br)+G(z)]\nonumber\\
&-g_1(\infty,z)G(z),
\label{g1j_final}
\end{align}
where
\begin{align}
F(\Delta\br)=&-\frac{\epsilon^2}{k_0}
\int\frac{d^2\bq}{(2\pi)^2}\gamma(\bq)\nonumber\\
&\times[2K(\bq)+2gI_0]\cos(\bq\cdot\Delta\br)
\label{F_expression}
\end{align}
and
\begin{align}
G(z)=-\frac{\epsilon^2}{k_0}
\int\!\frac{d^2\bq}{(2\pi)^2}2gI_0\gamma(\bq)[1-2\sin^2k(\bq)z].
\label{G_expression}
\end{align}
The function $F(\Delta\br)$ is short range, with a decay ranging over a few $\sigma$ and entirely controlled by the shape of the power spectrum $\gamma(\bq)$.
The correlator $g_1^j(\Delta\br,z)$ is thus short range as well -- see Eq. (\ref{g1j_final}) --  and nonuniversal.


\begin{thebibliography}{99}
\bibitem{footnote1}
Since the system is translation invariant on average, both the speckle correlation and the coherence function are independent of $\br_\perp$.
\end{thebibliography}
\end{document}